\documentclass[aps,twocolumn,prl,showpacs,amsmath,amssymb,10pt]{revtex4-2}
\usepackage{graphicx}
\usepackage{scrextend}
\usepackage{manfnt,pifont}
\usepackage{hyperref}
\usepackage{url}
\usepackage{lipsum}
\usepackage[font=small]{subcaption}
\usepackage[font=small]{caption}

\usepackage[normalem]{ulem}

\usepackage{url}
\usepackage[outline]{contour}
\usepackage{textgreek}

\usepackage{tikz}
\usetikzlibrary{shapes}

\usepackage{CJK}
\usepackage{enumitem}
\setlist[itemize]{leftmargin=*}

\newcommand\wh\widehat

\makeatletter
\DeclareFontFamily{OMX}{MnSymbolE}{}
\DeclareSymbolFont{MnLargeSymbols}{OMX}{MnSymbolE}{m}{n}
\SetSymbolFont{MnLargeSymbols}{bold}{OMX}{MnSymbolE}{b}{n}
\DeclareFontShape{OMX}{MnSymbolE}{m}{n}{
    <-6>  MnSymbolE5
   <6-7>  MnSymbolE6
   <7-8>  MnSymbolE7
   <8-9>  MnSymbolE8
   <9-10> MnSymbolE9
  <10-12> MnSymbolE10
  <12->   MnSymbolE12
}{}
\DeclareFontShape{OMX}{MnSymbolE}{b}{n}{
    <-6>  MnSymbolE-Bold5
   <6-7>  MnSymbolE-Bold6
   <7-8>  MnSymbolE-Bold7
   <8-9>  MnSymbolE-Bold8
   <9-10> MnSymbolE-Bold9
  <10-12> MnSymbolE-Bold10
  <12->   MnSymbolE-Bold12
}{}

\let\llangle\@undefined
\let\rrangle\@undefined
\DeclareMathDelimiter{\llangle}{\mathopen}%
                     {MnLargeSymbols}{'164}{MnLargeSymbols}{'164}
\DeclareMathDelimiter{\rrangle}{\mathclose}%
                     {MnLargeSymbols}{'171}{MnLargeSymbols}{'171}
\makeatother

\usepackage{eso-pic}
\newcommand{\reportnum}[2]{
  \AddToShipoutPictureBG*{%
    \AtPageUpperLeft{%
      \hspace{0.75\paperwidth}%
      \raisebox{#1\baselineskip}{%
        \makebox[0pt][l]{\textnormal{#2}}
  }}}%
}

\begin{document}

\begin{CJK*}{UTF8}{GB}
\CJKfamily{gbsn}

\reportnum{-3}{USTC-ICTS/PCFT-24-16}

\title{Unitarity Method for Holographic Defects}

\author{Junding Chen (陈俊定)$^{a,b}$}
\author{Aleix Gimenez-Grau$^{c}$}
\author{Hynek Paul$^{d}$}
\author{Xinan Zhou (周稀楠)$^{e,f}$}
\affiliation{$^{a}$CAS Key Laboratory of Theoretical Physics, Institute of Theoretical Physics, Chinese Academy of Sciences, Beijing 100190, China}
\affiliation{$^b$School of Physical Sciences, University of Chinese Academy of Sciences, No.19A Yuquan Road, Beijing 100049, China}
\affiliation{$^{c}$Institut des Hautes \'Etudes Scientifiques, 91440 Bures-sur-Yvette, France}
\affiliation{$^{d}$Instituut voor Theoretische Fysica, KU Leuven, Celestijnenlaan 200D, B-3001 Leuven, Belgium}
\affiliation{$^{e}$Kavli Institute for Theoretical Sciences, University of Chinese Academy of Sciences, Beijing 100190, China}
\affiliation{$^{f}$Peng Huanwu Center for Fundamental Theory, Hefei, Anhui 230026, China}

\begin{abstract}
 We initiate the study of loop-level holographic correlators in the presence of defects. We present a unitarity method which constructs loop corrections from lower order data. As an example, we apply this method to 6d $\mathcal{N}=(2,0)$ theories with $\frac{1}{2}$-BPS surface defects and report the first holographic two-point function at one loop.

\end{abstract}

	\maketitle
\end{CJK*}

\noindent {\bf Introduction.} Defects are extended objects which lead to an interesting enrichment of theories. There is a wide range of motivations to consider defects, both experimentally (as impurities, domain walls, boundary effects, etc.) and formally (e.g., Wilson loops, D-branes and symmetry generators). The AdS/CFT correspondence offers yet another reason to study them. In its most studied form, the correspondence is a duality between weakly coupled AdS gravity and a boundary CFT with many degrees of freedom (parameterized by a large central charge $c$). CFT correlators are holographically mapped to scattering amplitudes in AdS where the $1/c$ expansion amounts to a loop expansion. With the addition of defects, the theory admits new observables, namely correlators of operators inserted away from the defect, which become AdS {\it form factors} of particles scattering off extended objects. While holographic correlators are notoriously difficult to compute by Witten diagrams (Feynman diagrams in AdS), the recent analytic bootstrap program, initiated in \cite{Rastelli:2016nze,Rastelli:2017udc}, allowed us to go way beyond and revealed remarkable simplicity and hidden structures (see \cite{Bissi:2022mrs} for a review). The most remarkable lesson from this program is that these correlators are largely fixed by symmetries and consistency conditions. It is natural to ask  if we can extend the success to include defects and uncover interesting new physics. Unfortunately, not much is known for holographic defect correlators. So far all the bootstrap analyses have focused on the first nontrivial order in $1/c$ corresponding to tree-level processes \cite{Barrat:2022psm,Bianchi:2022ppi,Barrat:2021yvp,Meneghelli:2022gps,Gimenez-Grau:2023fcy,Chen:2023yvw}, with no results at higher orders \footnote{There are loop-level results when all operators are inserted on the defect \cite{Ferrero:2021bsb,Ferrero:2023gnu}. However, these are kinematically the same as defect-free correlators and do not correspond to AdS form factors. }\footnote{Using supersymmetric localization, one can compute integrated two-point functions, see \cite{Pufu:2023vwo,Billo:2023ncz,Dempsey:2024vkf,Billo:2024kri} for recent progress.}.

\begin{figure}
    \centering
   \includegraphics[width=0.5\textwidth]{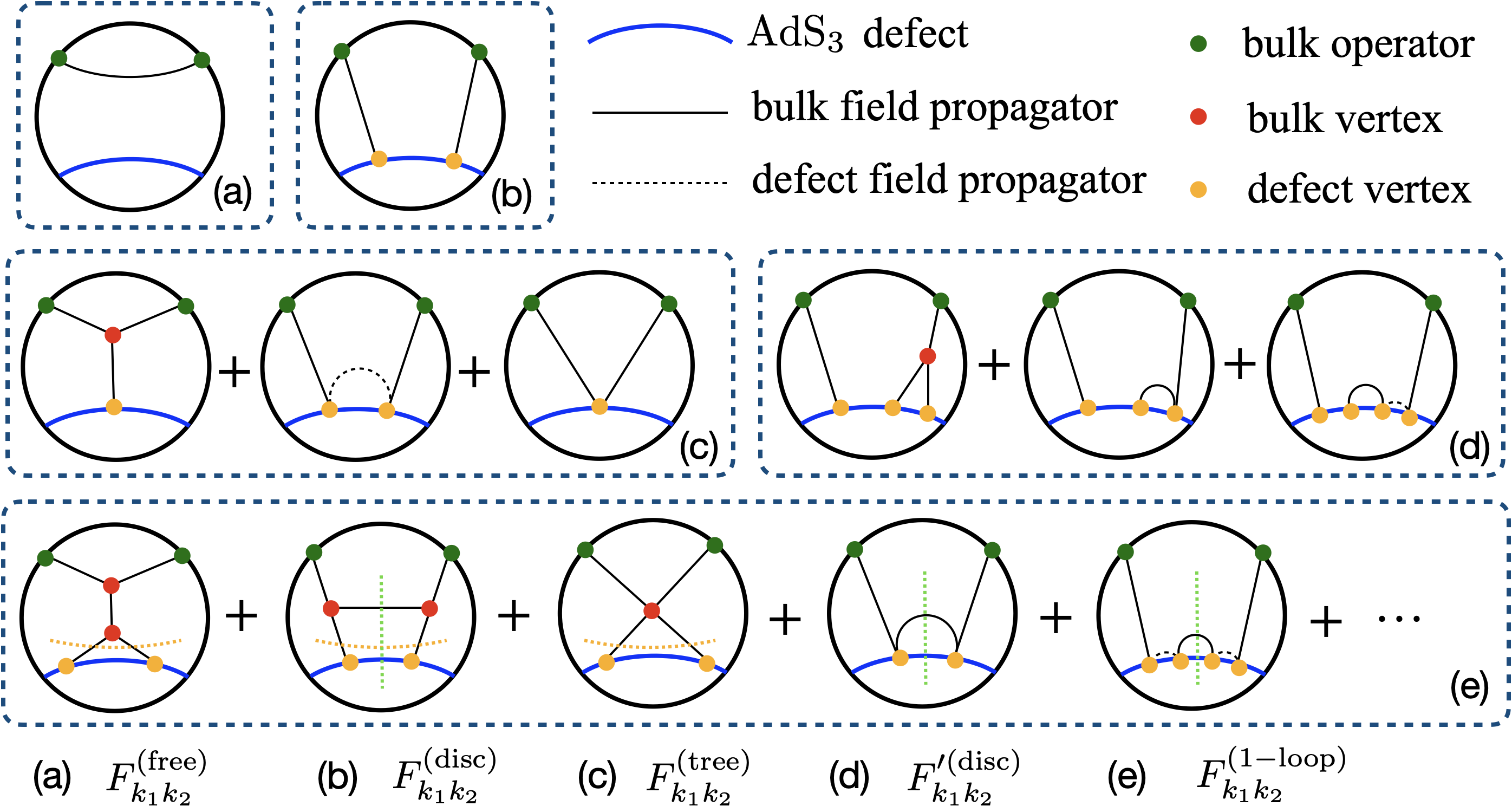}
    \caption{Witten diagrams in the first few orders of the large $N$ expansion.}
    \label{fig:Wittendiagrams}
\end{figure}

In this paper, we initiate the study of loop-level defect correlators. We focus on two-point functions and consider holographic theories where defects are dual to AdS branes (with possible internal factors). This includes an array of setups where the defect AdS branes can be identified with string worldsheets and probe D-branes. For concreteness, we base our discussion in 6d $\mathcal{N}=(2,0)$ theories with $\frac{1}{2}$-BPS surface defects. However, this sacrifices no generality and generalizing to other setups is straightforward. Constructed from $N\gg1$ M5 branes and $Q$ probe M2 branes, the theory is dual to  11D supergravity in AdS$_7\times $S$^4$ coupled to an AdS$_3$ brane carrying localized defect degrees of freedom. Two-point functions of bulk KK modes admit the expansion \cite{Meneghelli:2022gps}\footnote{Here we do not consider M-theory corrections.}
\begin{equation}\label{Fk1k2exp}
\begin{split}
\llangle \mathcal{S}_{k_1}\mathcal{S}_{k_2}\rrangle={}&F_{k_1k_2}^{({\rm free})}+\frac{d^2}{c} F_{k_1k_2}^{({\rm disc})}+\frac{d}{c} F_{k_1k_2}^{({\rm tree})}\\
{}&+\frac{d^3}{c^2} F_{k_1k_2}'^{({\rm disc})}+\frac{d^2}{c^2} F_{k_1k_2}^{({\rm 1-loop})}+\ldots\;,
\end{split}
\end{equation}
where $c\sim N^3$, $d\sim QN$ are bulk and defect anomaly coefficients \cite{Henningson:1998gx,Intriligator:2000eq,Tseytlin:2000sf,Beccaria:2014qea,brv14,Estes:2018tnu,Jensen:2018rxu,Drukker:2020swu} and the $i$-th term in the expansion is of order $\mathcal{O}(N^{-i+1})$. As illustrated in Fig. \ref{fig:Wittendiagrams}, the leading term $F_{k_1k_2}^{({\rm free})}$ is given by a free AdS propagator. The disconnected contributions $F_{k_1k_2}^{({\rm disc})}$, $F_{k_1k_2}'^{({\rm disc})}$ also trivially factorize into one-point functions. The first nontrivial term is the tree-level $F_{k_1k_2}^{({\rm tree})}$ and was computed in \cite{Meneghelli:2022gps} (and in full generality in \cite{Chen:2023yvw}). Here our target is the one-loop contribution $F_{k_1k_2}^{({\rm 1-loop})}$.

\begin{figure}
    \centering
   \includegraphics[width=0.48\textwidth]{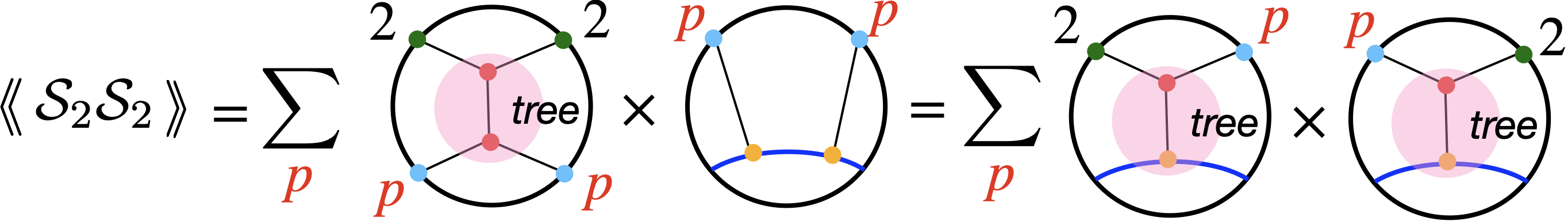}
    \caption{One-loop correlator from tree-level correlators.}
    \label{fig:unitarity}
\end{figure}

Our strategy for computing these loop corrections is by ``gluing'' together tree-level correlators. This can be partially illustrated by the underlying Witten diagrams. Cutting along the yellow or green dashed lines in Fig.~\ref{fig:Wittendiagrams}(e) leads to either defect-free tree-level four-point diagrams and disconnected defect two-point functions, or tree-level defect two-point diagrams. We hasten to add that this intuitive picture is too schematic and even slightly misleading because most diagrams {\it cannot} be cut in two ways. Instead, as illustrated in Fig. \ref{fig:unitarity}, our method leverages that the one-loop correlator can be obtained from gluing in {\it both} ways. In fact, the consistency of the two gluing operations is an important check of our method and is a manifestation of equivalence of OPEs in different channels. We also need to sum over all the KK modes in the glued propagators, which is correctly reflected in the diagrammatic picture. Our gluing strategy is reminiscent of the flat-space unitarity method (see e.g., \cite{Elvang:2015rqa} for an introduction) which has been introduced to defect-free holographic CFTs for one-loop four-point functions in \cite{Aharony:2016dwx}. The goal of this paper is to extend it to include holographic defects and turn this imprecise gluing picture into a concrete algorithm for computing loop-level defect correlators. We will achieve this by first constructing certain logarithmic singularities from tree-level data and then completing them into a full correlator in Mellin space. As an explicit example, we will compute the one-loop  two-point function of the lowest KK mode. 

\vspace{0.4cm}

\noindent {\bf Kinematics.} We focus on $\frac{1}{2}$-BPS operators $\mathcal{S}_k(x,u)\equiv\mathcal{S}_k^{I_1\ldots I_k}u_{I_1}\ldots u_{I_k}$, dual to bulk supergravitons ($k=2$) and higher KK modes ($k\geq3$). Here $I=1,\ldots,5$ are indices of the R-symmetry group $SO(5)$ which is the isometry of S$^4$. We introduce $SO(5)$ null vectors  $u^I$ to keep track of the R-symmetry information. The $\frac{1}{2}$-BPS defect divides the coordinates into a parallel part $x^{a=1,2}$ and a transverse part $x^{i=3,\ldots,6}$, and preserves $[OSp(4^*|2)]^2\subset OSp(8^*|4)$ superconformal symmetry. Defect two-point functions can be written as 
\begin{equation}
\llangle \mathcal{S}_{k_1}\mathcal{S}_{k_2}\rrangle=\frac{(u_1\cdot \theta)^{k_1}(u_2\cdot \theta)^{k_2}}{|x_1^i|^{2k_1}|x_2^i|^{2k_2}}\mathcal{F}_{k_1k_2}(z,\bar{z},\sigma)\;,
\end{equation}
where $z$, $\bar{z}$ are the conformal cross-ratios
\begin{equation}
\frac{x_{12}^2}{|x_1^i||x_2^i|}=\frac{(1-z)(1-\bar{z})}{\sqrt{z\bar{z}}}\;,\quad \frac{x_1^jx_2^j}{|x_1^i||x_2^i|}=\frac{z+\bar{z}}{\sqrt{z\bar{z}}}\;,
\end{equation}
and $\sigma$ is the R-symmetry cross-ratio
\begin{equation}
\sigma=\frac{u_1\cdot u_2}{(u_1\cdot\theta)(u_2\cdot\theta)}=-\frac{(1-\omega)^2}{2\omega}\;.
\end{equation}
We have introduced a fixed unit vector $\theta$ with $\theta\cdot\theta=1$ to encode the R-symmetry breaking from $SO(5)$ to $SO(4)$ by the defect which sits at a point on S$^4$. The correlator $\mathcal{F}_{k_1k_2}$ is a polynomial in $\sigma$ with degree $k_{\rm m}=\min\{k_1,k_2\}$. Superconformal Ward identities further constrain $\mathcal{F}_{k_1k_2}$ to take the form \cite{Meneghelli:2022gps}
\begin{equation}
\nonumber\mathcal{F}_{k_1k_2}(z,\bar{z},\omega)=\mathcal{F}_{{\rm prot},k_1k_2}(z,\bar{z},\omega)+\mathrm{R}\,\mathcal{H}_{k_1k_2}(z,\bar{z},\omega)\;,
\end{equation}
where $\mathcal{F}_{{\rm prot},k_1k_2}$ is a protected piece and $\mathrm{R}=(z\bar{z})^{-1}(z-\omega)(\bar{z}-\omega)(z-\omega^{-1})(\bar{z}-\omega^{-1})$ is a factor determined by superconformal symmetry. Finally, all dynamical information is encoded in the reduced correlator $\mathcal{H}_{k_1k_2}$ which is of lower degree $k_{\rm m}-2$ in $\sigma$. 

To make its analytic structure manifest, it is useful to use the Mellin representation of defect correlators \cite{Rastelli:2017ecj,Goncalves:2018fwx}, which in many ways is reminiscent of flat-space momentum space. Defining the combinations
\begin{equation}
\mathcal{B}=\frac{(1-z)(1-\bar{z})}{z+\bar{z}}\;,\quad \mathcal{D}=\frac{\sqrt{z\bar{z}}}{z+\bar{z}}\;,
\end{equation}
which go to zero in the bulk and defect channel OPE respectively,  we can write the reduced correlator as a two-fold inverse Mellin transformation
\begin{equation}\label{defMellin}
\mathcal{H}_{k_1k_2}=\int \frac{d \delta \, d\gamma}{(2\pi i)^2}  
\mathcal{B}^{-\delta} \mathcal{D}^\gamma  \widetilde{\mathcal{M}}_{k_1k_2}(\delta,\gamma)\widetilde{\Gamma}_{k_1k_2}(\delta,\gamma)\;,
\end{equation}
with $\widetilde{\Gamma}_{k_1k_2}(\delta,\gamma)= \Gamma(\delta)\Gamma(\gamma-\delta)\prod_{i=1}^2\Gamma\!\left(\frac{2k_i+2-\gamma}{2}\right)$. All information is encoded in the reduced Mellin amplitude $\widetilde{\mathcal{M}}_{k_1k_2}$. The full Mellin amplitude $\mathcal{M}_{k_1k_2}$ can be similarly defined from $\mathcal{F}_{k_1k_2}$ and is related to $\widetilde{\mathcal{M}}_{k_1k_2}$ by a difference operator \cite{Chen:2023yvw}, which however will not be needed in this paper.

\vspace{0.4cm}

\noindent {\bf Logarithmic singularities.} The schematic gluing idea can be made precise in terms of the leading logarithmic singularities (LLS) as $\mathcal{B},\mathcal{D}\to 0$. These singularities are associated with the anomalous dimensions of unprotected operators and arise from individual conformal blocks (see the Supplemental Material for explicit expressions and our normalizations)   
\begin{equation}
\begin{split}
\nonumber {}&\widehat{f}_{\widehat{\Delta},s}\big|_{\mathcal{D}\to 0}\sim \mathcal{D}^{\widehat{\Delta}-s}(1+\ldots)=\mathcal{D}^{\widehat{\Delta}_0-s}(1+\widehat{\gamma}\log \mathcal{D}+\ldots)\;,\\
{}&f^{k_{12}}_{\Delta,\ell}\big|_{\mathcal{B}\to 0}\sim \mathcal{B}^{\frac{\Delta-\ell}{2}}(1+\ldots)=\mathcal{B}^{\frac{\Delta_0-\ell}{2}}(1+\frac{\gamma}{2}\log \mathcal{B}+\ldots)\;.
\end{split}
\end{equation}
Here $\widehat{f}_{\widehat{\Delta},s}$ and $f^{k_{12}}_{\Delta,\ell}$ are respectively the defect and bulk channel conformal blocks for exchanging operators with dimension $\widehat{\Delta}$, transverse spin $s$ and dimension $\Delta$, spin $\ell$. At large $N$, these operators have small anomalous dimensions $\widehat{\Delta}=\widehat{\Delta}_0+\widehat{\gamma}$, $\Delta=\Delta_0+\gamma$, and expanding in $\widehat{\gamma}$, $\gamma$ leads to logarithms in $\mathcal{D}$ and $\mathcal{B}$.  In the supergravity limit, all the long operators are of the double-trace type. In the bulk channel, these operators are schematically $:\square^m\mathcal{S}_{k_1}\partial^\ell\mathcal{S}_{k_2}:$. In the defect channel, they take the form $\square_\perp^m\partial_\perp^s\mathcal{S}_k$ for bulk operators $\mathcal{S}_k$ restricted to the defect. Note that in each channel, the double-trace operators are degenerate and generally will mix among themselves. In the following, we present the LLS structure at each order in (\ref{Fk1k2exp}). We write them in a conformal block decomposition, and document the explicit coefficients in the Supplemental Material. 
\begin{itemize}
\item {\it Free correlator:} The free correlator is given by
\begin{equation}
\label{freecorrelatorFkk}
\mathcal{F}_{kk}^{\rm (free)}=\left(\sigma \mathcal{D}^2\mathcal{B}^{-2}\right)^{k}\;,
\end{equation}
and contains no logarithmic singularities. The bulk channel decomposition contains only the identity operator while infinitely many defect double-trace operators appear in the defect channel decomposition   \begin{equation}
\label{Hfreedefectdec}
\mathcal{H}_{kk}^{\rm (free)}\big|_{\rm long}\supset\!\sum_{m=-k+2,s=0}^{\infty}\langle (b^{(0)}_{k,m,s})^2\rangle \,\widehat{\bf f}_{2k+2m+s,s}\;,
\end{equation}
where $b^{(0)}_{k,m,s}$ are their coefficients in the defect channel OPE of $\mathcal{S}_k$. Here and below we denote with boldface ${\bf f}_{\Delta,\ell}$ and $\widehat{\bf f}_{\widehat \Delta, s}$ the superconformal blocks of the reduced correlator $\mathcal H$. The dimensions are shifted in $\widehat{\bf f}_{\widehat{\Delta},s}=\widehat{f}_{\widehat{\Delta}+2,s}$ due to superconformal symmetry (see (SM-7) in Supplemental Material).

\item {\it Disconnected level:} As the product of one-point functions, $\mathcal{F}_{k_1k_2}^{\rm (disc)}$ and $\mathcal{F}_{k_1k_2}'^{\rm (disc)}$ also have no logarithmic singularities. The leading disconnected piece is
\begin{equation}
\mathcal{F}_{k_1k_2}^{\rm (disc)}=\frac{\Gamma(k_1)\Gamma(k_2)}{\sqrt{2^{k_1+k_2}\Gamma(2 k_1-1)\Gamma(2 k_2-1)}}\;,
\end{equation}
and only exchanges the defect identity in the defect channel. In the bulk channel, we find infinitely many bulk double-trace operators
  \begin{equation}\label{Fdiscbulkdec}
\mathcal{F}_{kk}^{\rm (disc)}\big|_{\rm long}\supset\sum_{m,\ell=0,\, \text{even}}^{\infty} \langle\lambda^{(0)}_{k,m,\ell} a^{(0)}_{m,\ell}\rangle\, {\rm R}\,{\bf f}_{4k+2m+\ell, \ell}\;,
\end{equation}
where they have bulk OPE coefficients $\lambda^{(0)}_{k,m,\ell}$ with two $\mathcal{S}_k$ and one-point function coefficients $a^{(0)}_{m,\ell}$. The dimension shift in ${\bf f}_{\Delta, \ell}= \frac{1}{4}(\mathcal{D}\mathcal{B}^{-1})^{2k+2}\sigma^{k-2} f^{k_{12}=0}_{\Delta+4, \ell}$ is also because of supersymmetry (see (SM-11)). We focus on $k_1=k_2=k$ because only the operators $:\square^m\mathcal{S}_k\partial^\ell\mathcal{S}_k:$ mix with $:\square^n\mathcal{S}_2\partial^\ell\mathcal{S}_2:$ and will be relevant for computing $\llangle\mathcal{S}_2\mathcal{S}_2\rrangle_{\rm 1-loop}$. The coefficients $\lambda^{(0)}_{k,m,\ell}$ also appear in the defect-free disconnected four-point functions 
\begin{equation}
\mathcal{F}^{\rm 4pt, (disc)}_{kkkk}\big|_{\rm long}\supset\sum_{m,\ell}\langle(\lambda^{(0)}_{k,m,\ell})^2\rangle\, {\bf g}^{\rm 4pt}_{4k+2m+\ell,\ell}\;,
\end{equation}
where the explicit superconformal blocks can be found in \cite{Alday:2020tgi}. A similar analysis applies to $\mathcal{F}_{k_1k_2}'^{\rm (disc)}$, but it is of higher order in $1/N$ and is hence irrelevant for our one-loop analysis. Moreover, $\mathcal{F}_{22}'^{\rm (disc)}$ vanishes as shown in \cite{Meneghelli:2022gps}.

\item {\it Tree level:} Logarithmic singularities first appear in tree-level  correlators which were computed in full generality in \cite{Chen:2023yvw}. In the defect channel we can insert one power of defect anomalous dimensions $\widehat{\gamma}^{(1)}_{m,s}$ which are $\mathcal{O}(1/N^2)$. This gives a $\log \mathcal{D}$ LLS as $\mathcal{D}\to 0$ given by 
\begin{equation}
\label{H2ktreedefectdec}
\mathcal{H}^{\rm (tree)}_{2k}\big|_{\log\mathcal{D}}= \sum_{m,s=0}^{\infty} \langle b^{(0)}_{2,m,s} b^{(0)}_{k,m,s}\widehat{\gamma}^{(1)}_{m,s}\rangle \,\widehat{\bf f}_{2k+2m+s,s}\;.
\end{equation}
On the other hand, bulk operator anomalous dimensions are $\mathcal{O}(1/N^3)$ and hence cannot be inserted. Therefore, there is no $\log\mathcal{B}$ singularity. The bulk channel decomposition takes the same form as in (\ref{Fdiscbulkdec}) but with $a^{(0)}$ replaced by their tree-level corrections $a^{(1)}$.

\item {\it One-loop level:} At one loop, we focus on our target correlator with $k_1=k_2=2$. In the defect channel, the LLS is $\log^2\mathcal{D}$ and its coefficient is 
\begin{equation}
\label{H221loopdefectdec}
\mathcal{H}^{\rm (1-loop)}_{22}\big|_{\log^2\mathcal{D}}=\sum_{n,s=0}^{\infty} \frac{1}{2} \langle (b^{(0)}_{2,n,s})^2(\widehat{\gamma}_{n,s}^{(1)})^2\rangle \,\widehat{\bf f}_{4+2n+s,s}\;.
\end{equation}
In the bulk channel, we can insert one power of the bulk anomalous dimension and it leads to a $\log\mathcal{B}$ singularity 
\begin{equation}
\label{H221loopbulkdec}
\mathcal{H}^{\rm (1-loop)}_{22}\big|_{\log\mathcal{B}}=\sum_{n,\ell=0\,\text{even}}^{\infty}\frac{1}{2} \langle\lambda^{(0)}_{2,n,\ell} a^{(0)}_{n,\ell} \gamma^{(1)}_{n,\ell}\rangle \,{\bf f}_{8+2n+\ell,\ell}\;.
\end{equation}
The bulk anomalous dimensions $ \gamma^{(1)}_{n,\ell}$ also appear in the defect-free tree-level four-point functions \cite{Rastelli:2017ymc,Zhou:2017zaw,Alday:2020lbp} as
\begin{equation}
\nonumber \mathcal{F}^{\rm 4pt,(tree)}_{22kk}\big|_{\text{log part}}\supset \sum_{m,\ell}\frac{1}{2} \langle\lambda^{(0)}_{2,m,\ell} \lambda^{(0)}_{k,m,\ell}\gamma^{(1)}_{m,\ell}\rangle\,{\bf g}_{4k+2m+\ell,\ell}^{\rm 4pt}\;,
\end{equation}
and the OPE coefficients were extracted in \cite{Alday:2020tgi}.
\end{itemize}
Clearly, the LLS at one loop depends only on CFT data from lower orders. A complication that prevents us from simply multiplying and dividing the decomposition coefficients to get the one-loop ones is operator mixing as indicated by the averages $\langle \ldots \rangle$. This difficulty can be overcome by  considering correlators of different KK modes which gives enough equations to unmix the data. However, it turns out that performing the explicit unmixing is not needed for computing  $\llangle \mathcal{S}_2\mathcal{S}_2\rrangle_{\rm 1-loop}$, and will therefore be presented elsewhere \cite{defectlong}. As we will show later, we only need to assemble the decomposition coefficients of the two-point functions $\llangle \mathcal{S}_p\mathcal{S}_p\rrangle_{\rm free}$,  $\llangle \mathcal{S}_p\mathcal{S}_p\rrangle_{\rm disc}$,  $\llangle \mathcal{S}_2\mathcal{S}_p\rrangle_{\rm tree}$, and the four-point functions $\langle \mathcal{S}_p\mathcal{S}_p\mathcal{S}_p\mathcal{S}_p\rangle_{\rm disc} $, $\langle \mathcal{S}_2\mathcal{S}_2\mathcal{S}_p\mathcal{S}_p\rangle_{\rm tree} $ for all values of $p$.

\vspace{0.4cm}

\noindent {\bf Unitarity method for defects.} After obtaining the one-loop LLS from the tree-level data (to be further detailed in a moment), the next step is to complete them into the full correlator. In principle, this can be achieved by using the defect dispersion relations \cite{Barrat:2022psm,Bianchi:2022ppi}. However, the resulting integrals are difficult to evaluate. We will instead propose an algorithm in Mellin space to exploit its simpler analytic structure. A similar algorithm has been used for one-loop four-point Mellin amplitudes without defects in \cite{Alday:2018kkw,Alday:2019nin,Alday:2021ajh,Huang:2023ppy}. It should be emphasized that the algorithm is not restricted to $\llangle \mathcal{S}_2\mathcal{S}_2\rrangle$ and generalizes straightforwardly to correlators of higher KK modes. 

The one-loop correlator can be written as 
\begin{equation}
\begin{split}
\nonumber \mathcal{F}_{22}^{({\rm 1-loop})}={}&F_{1,2}\log \mathcal{B} \log^2\mathcal{D} +F_{1,1}\log \mathcal{B} \log\mathcal{D} +F_{1,0}\log\mathcal{B}\\
{}&+F_{0,2}\log^2\mathcal{D} +F_{0,1}\log\mathcal{D} +F_{0,0}\;,
\end{split}
\end{equation}
where $F_{i,j}$ are regular as $\mathcal{B}, \mathcal{D}\to 0$. The coefficients $F_{1,j}$ and  $F_{i,2}$ constitute the bulk and defect channel LLS respectively. The common coefficient $F_{1,2}$ can be expanded as 
\begin{equation}\label{F12exp}
F_{1,2}=\sum_{m,n=0}^\infty d_{mn}\mathcal{B}^n\mathcal{D}^{6+2m}\;.
\end{equation}
Reproducing it in Mellin space requires the integrand in eq.~(\ref{defMellin}) to have double poles at $\delta=-n$ and triple poles at $\gamma=6+2m$. Since the Gamma factor already contains poles at these locations, the reduced Mellin amplitude $\widetilde{\mathcal{M}}^{(\rm 1-loop)}_{22}$ must contain simple poles in $\delta$ and $\gamma$. We will make the following seemingly radical ansatz:  The reduced Mellin amplitude contains only {\it simultaneous} poles 
\begin{equation}\label{Mansatz}
\widetilde{\mathcal{M}}^{(\rm 1-loop)}_{22}=\sum_{m,n=0}^\infty \frac{c_{mn}}{(\delta+n)(\gamma-6-2m)}\;,
\end{equation}
with {\it constant} numerators $c_{mn}$. Here the range of poles is justified by the fact that only long operators contribute at one loop. However, that the $c_{mn}$ are constants is a genuine assumption which needs to be checked by calculations. To this end, we proceed as follows:
\begin{enumerate}
\item We take residues and focus on the $\log^2\mathcal{D}\log\mathcal{B}$ coefficient. By matching with (\ref{F12exp}), we obtain $c_{mn}$. 
\item With these $c_{mn}$, we take residues in $\gamma$ and focus on the $\log^2 \mathcal{D}$ coefficient (with full $\mathcal{B}$ dependence). We find that all $F_{i,2}$ are correctly reproduced, implying that there cannot be additional single poles in $\gamma$. 
\item Similarly, taking residues in $\delta$ and focusing on the $\log\mathcal{B}$ coefficients, we find that all $F_{1,j}$ are reproduced. This rules out additional single poles in $\delta$. 
\end{enumerate}
This procedure fixes the one-loop reduced Mellin amplitude up to regular terms. Such polynomial ambiguities correspond to contact diagrams, which can be interpreted as counter terms for the UV divergence \footnote{This is the case for four-point functions and can be proven by using the flat-space limit.}. In the following, we carry out the above algorithm and show how the tree-level data precisely enters at one loop. 

\vspace{0.4cm}

\noindent {\bf Defect channel.} We first consider defect double-trace operators $\{\square_\perp^{L-2}\partial_\perp^s\mathcal{S}_2,\ldots, \partial_\perp^s\mathcal{S}_{L}\}$ for $L=2,3,\ldots$. These operators have the same defect twist $2L$ at infinite $N$ and  develop anomalous dimensions at $\mathcal{O}(N^{-2})$. To keep track of mixing, we organize various CFT data in matrices and choose a basis $\{\widehat{O}_1,\ldots,\widehat{O}_{L-1}\}$ where  anomalous dimensions are diagonalized. We can organize the defect OPE coefficients as 
\begin{equation}
\widehat{\bf B}=\left(\begin{array}{cccc}b^{(0)}_{2\widehat{O}_1} & b^{(0)}_{2\widehat{O}_2} & \ldots & b^{(0)}_{2\widehat{O}_{L-1}} \\b^{(0)}_{3\widehat{O}_1} & b^{(0)}_{3\widehat{O}_2} & \ldots & b^{(0)}_{3\widehat{O}_{L-1}} \\\ldots & \ldots & \ldots & \ldots \\b^{(0)}_{L\widehat{O}_1} & b^{(0)}_{L\widehat{O}_2} & \ldots & b^{(0)}_{L\widehat{O}_{L-1}}\end{array}\right).
\end{equation} 
For the free correlator, we then have 
\begin{equation}
\nonumber \widehat{\bf B}\widehat{\bf B}^{T}=\widehat{\bf N}={\rm diag}\{\llangle \mathcal{S}_2\mathcal{S}_2\rrangle,\ldots,\llangle \mathcal{S}_L\mathcal{S}_L\rrangle\}_{\rm free}\;.
\end{equation}
Defining the matrix of tree-level anomalous dimensions $\widehat{\bf \Gamma}={\rm diag}\{\widehat{\gamma}^{(1)}_1,\ldots, \widehat{\gamma}^{(1)}_{L-1}\}$, we  have 
\begin{equation}
\nonumber\widehat{\bf \Omega}=\widehat{\bf B}\widehat{\bf \Gamma}\widehat{\bf B}^{T}=\left(\begin{array}{ccc}\llangle \mathcal{S}_2\mathcal{S}_2\rrangle & \ldots & \llangle \mathcal{S}_2\mathcal{S}_L\rrangle\\\ldots & \ldots & \ldots \\\llangle \mathcal{S}_L\mathcal{S}_2\rrangle & \ldots & \llangle \mathcal{S}_L\mathcal{S}_L\rrangle
\end{array}\right)^{\rm tree}_{\log\mathcal{D}}.
\end{equation} 
At one loop, the $\log^2\mathcal{D}$ coefficients (\ref{H221loopdefectdec}) are given by the top-left component of the matrix
\begin{equation}
\widehat{\bf B}(\widehat{\bf \Gamma})^2\widehat{\bf B}^{T}=\widehat{\bf \Omega}\,\widehat{\bf N}^{-1}\widehat{\bf \Omega}\;,
\end{equation}
implying that we only need $\llangle \mathcal{S}_p\mathcal{S}_p\rrangle_{\rm free}$ and $\llangle \mathcal{S}_2\mathcal{S}_p\rrangle_{\rm tree}$. The coefficients are given in terms of matrix elements as
\begin{equation}
\frac{1}{2}\langle (b^{(0)}_{2,n,s})^2 (\widehat{\gamma}^{(1)}_{n,s})^2\rangle=\sum_{k=2}^{L}\frac{\langle b^{(0)}_{2,m_{\rm d},s}b^{(0)}_{k,m_{\rm d},s} \widehat{\gamma}_\textsubscript{\textit{$m_{\rm d}$,s}}^{(1)}\rangle^2}{2 \,\langle (b^{(0)}_{k,m_{\rm d},s})^2\rangle}\;,
\end{equation}
with $m_{\rm d}=L-k$. We can resum the conformal blocks and  obtain the defect channel leading logarithmic singularity
\begin{equation}\label{LLSdefect}
(F_{1,2}\log\mathcal{B}+F_{0,2})\log ^2 \mathcal{D}\;,
\end{equation} in a power expansion of small $\mathcal{D}$ (for the first few orders see (SM-17) and (SM-18) in the Supplemental Material). From the coefficient function $F_{1,2}$ we can extract the simultaneous pole coefficients for fixed $m$ and arbitrary $n$, e.g.
\begin{equation}\label{cmnexamplen}
\begin{split}
{}&c_{0n}=\tfrac{9 \left(n^4+10 n^3+35 n^2+50 n+48\right)}{4 (n+1) (n+2) (n+3) (n+4) (n+5)}\;,\\
{}&c_{1n}=\tfrac{9 \left(5 n^6+81 n^5+517 n^4+1655 n^3+2814 n^2-464 n+1536\right)}{4 (n+1) (n+2) (n+3) (n+4) (n+5) (n+6) (n+7)}\;.
\end{split}
\end{equation}
We can then check that $F_{0,2}$ is also correctly reproduced, ruling out additional single poles in $\gamma$. 

\vspace{0.4cm}

\noindent {\bf Bulk channel.} Similarly, the bulk channel double-trace operators $\{\mathcal{S}_2\square^{2M-4}\partial^\ell\mathcal{S}_2,\ldots,\mathcal{S}_M\partial^\ell\mathcal{S}_M\}$, $M=2,3,\ldots$, all have degenerate twist $4M$ at infinite $N$ and develop anomalous dimensions at $\mathcal{O}(1/N^3)$. In the diagonal basis of the anomalous dimensions $\{O_1,\ldots,O_{M-1}\}$, we organize the three-point function coefficients into the matrix 
\begin{equation}
\mathbf{\Lambda}=\left(\begin{array}{cccc}\lambda^{(0)}_{22O_1} & \lambda^{(0)}_{22O_2} & \ldots & \lambda^{(0)}_{22O_{M-1}} \\\lambda^{(0)}_{33O_1} & \lambda^{(0)}_{33O_2} & \ldots & \lambda^{(0)}_{33O_{M-1}} \\\ldots & \ldots & \ldots & \ldots \\ \lambda^{(0)}_{MMO_1} & \lambda^{(0)}_{MMO_2} & \ldots & \lambda^{(0)}_{MMO_{M-1}}\end{array}\right),
\end{equation} 
such that for disconnected four-point functions we have
\begin{equation}
\nonumber {\bf \Lambda}{\bf \Lambda}^{T}={\bf N}={\rm diag}\{\langle \mathcal{S}_2\mathcal{S}_2\mathcal{S}_2\mathcal{S}_2\rangle,\ldots,\langle \mathcal{S}_M\mathcal{S}_M\mathcal{S}_M\mathcal{S}_M\rangle\}_{\rm disc}\;.
\end{equation}
Let us also define ${\bf\Gamma}={\rm diag}\{\gamma^{(1)}_1,\ldots,\gamma^{(1)}_{M-1}\}$ and a vector ${\bf A}=(a^{(0)}_1,\ldots,a^{(0)}_{M-1})^T$. At tree-level, we then have
\begin{equation}
\nonumber {\bf \Omega}={\bf \Lambda}{\bf \Gamma}{\bf \Lambda}^{T}=\left(\begin{array}{ccc}\langle \mathcal{S}_2\mathcal{S}_2\mathcal{S}_2\mathcal{S}_2\rangle & \ldots & \langle \mathcal{S}_2\mathcal{S}_2\mathcal{S}_M\mathcal{S}_M\rangle\\\ldots & \ldots & \ldots \\\langle \mathcal{S}_M\mathcal{S}_M\mathcal{S}_2\mathcal{S}_2\rangle & \ldots & \langle \mathcal{S}_M\mathcal{S}_M\mathcal{S}_M\mathcal{S}_M\rangle
\end{array}\right)^{\rm tree}_{\rm log}.
\end{equation} 
The one-loop level coefficient is given by the first component of the vector
\begin{equation}
{\bf \Lambda}{\bf \Gamma}{\bf A}={\bf \Omega}\,{\bf N}^{-1} ({\bf \Lambda} {\bf A})\;,
\end{equation}
where ${\bf \Lambda} {\bf A}=(\llangle\mathcal{S}_2\mathcal{S}_2\rrangle,\ldots \llangle\mathcal{S}_M\mathcal{S}_M\rrangle)^T_{\rm disc}$. Therefore, we only need the decomposition coefficients of $\llangle \mathcal{S}_p\mathcal{S}_p\rrangle_{\rm disc}$, $\langle \mathcal{S}_p\mathcal{S}_p\mathcal{S}_p\mathcal{S}_p\rangle_{\rm disc}$, $\langle \mathcal{S}_2\mathcal{S}_2\mathcal{S}_p\mathcal{S}_p\rangle_{\rm tree}$. In terms of matrix elements, we have 
\begin{equation}
\begin{aligned}
\label{oneloopOPEcoeffs}
&\frac{1}{2}\langle \lambda^{(0)}_{2,n,\ell}a^{(0)}_{n,\ell}\gamma^{(1)}_{n,\ell}\rangle\\&\quad\quad=\sum_{k=2}^{M}\frac{\langle\lambda^{(0)}_{k,m_{\rm b},\ell}a^{(0)}_{m_{\rm b},\ell}\rangle \langle\lambda^{(0)}_{2,m_{\rm b},\ell} \lambda^{(0)}_{k,m_{\rm b},\ell}\gamma^{(1)}_{m_{\rm b},\ell}\rangle}{2\,\langle(\lambda^{(0)}_{k,m_{\rm b},\ell})^2\rangle}\;,
\end{aligned}
\end{equation}
where $m_{\rm b}=2M-2k$. The leading logarithmic singularity can then be computed by resumming the bulk channel conformal blocks and gives  
\begin{equation}\label{LLSbulk}
(F_{1,2}\log^2\mathcal{D}+F_{1,1}\log\mathcal{D}+F_{1,0})\log\mathcal{B}\;,
\end{equation}
in a small $\mathcal{B}$ expansion (see eqs. (SM-19)--(SM-21)). The matching of $F_{1,2}$ from (\ref{LLSdefect}) and (\ref{LLSbulk}) is a nontrivial check of the consistency of our approach. We can extract from it the coefficients $c_{mn}$ for fixed $n$, e.g.
\begin{equation}\label{cmnexamplem}
\begin{split}
{}&c_{m0}=2\,c_{m1}=\tfrac{9 \sqrt{\pi } \left(35 m^2+125 m+96\right) \Gamma (m+1)}{512 \Gamma \left(m+\frac{7}{2}\right)}\;,\\
{}&c_{m2}=\tfrac{9 \sqrt{\pi } \left(329 m^3+1865 m^2+2830 m+1024\right) \Gamma (m+1)}{4096 \Gamma
   \left(m+\frac{9}{2}\right)}\;,
\end{split}
\end{equation}
which match (\ref{cmnexamplen}) for overlapping values of $(m,n)$ as expected. Furthermore, we can also check that the subleading logarithmic terms $F_{1,1}$ and $F_{1,0}$ are correctly reproduced, which forbids additional single poles in $\delta$. 

\vspace{0.4cm}

\noindent {\bf One-loop amplitude.} We can perform the calculations in the bulk and defect channels to very high orders and verify the absence of single poles. This also gives a large number of coefficients $c_{mn}$ for either fixed $m$ or $n$. From these examples, we managed to find the following closed form expression for the general $c_{mn}$ coefficient \footnote{This expression is superficially singular for $(m,n)$ at $(0, 0), (0, 1), (0, 2), (1, 0)$. The correct answer is given by first evaluating $m$ and then $n$.}
\begin{equation}
c_{mn}=p_{m}H_0+q_{m,n}H_1+r_{m,n}H_2+s_{m}H_4\;,
\end{equation}
where
\begin{equation}
\begin{aligned}
&H_a(m,n)=\bigg[   \frac{\sqrt{\pi }4^{ m}\Gamma (m+n+3)}{\Gamma(\tfrac{-2m-3}{2}) \Gamma (2 m+n+6)} \\&\quad~\times{}_3 F_2\left(\left.\begin{gathered}
-m-2\,,\tfrac{-2m-n-5}{2}\,, \tfrac{-2m-n-4}{2} \\
\tfrac{-2m-3}{2}\,, -m-n-2
\end{gathered} \right\rvert\, 1\right)\Bigg]\Bigg|_{m\to m-a},
\end{aligned}
\end{equation}
and
\begin{equation}
\begin{aligned}
p_m=&\;3 (m+1)^2 (m+2)^2\;,\\q_{m,n}=&\;7 m^2 n^2+28 m^3 n-21 m^2 n+64 m^4\\&+134 m^3+158 m^2-14 m n^2+6 m n\\&+250 m+19 n^2+95 n+162\;,\\r_{m,n}=&-7 m^2 n^2-28 m^3 n+35 m^2 n-102 m^4\\&-82 m^3-158 m^2+14 m n^2-34 m n\\&-230 m-19 n^2-57 n-100\;,\\s_m=&\;35 (m-1)^2 m^2\;.
\end{aligned}
\end{equation}
This constitutes the main result of this paper. A curious feature is the appearance of hypergeometric functions, as opposed to simple rational functions in the four-point case \cite{Alday:2018kkw,Alday:2019nin,Alday:2021ajh,Huang:2023ppy}. 

\vspace{0.4cm}

\noindent {\bf Discussion.} In this paper we described a unitarity method which computes loop corrections to holographic defect correlators from lower order data. We also reported the first one-loop  defect two-point function. A more detailed exposition will be presented in \cite{defectlong}. There are many interesting future research avenues.

First, an immediate generalization is to perform a systematic analysis for one-loop correlators of higher KK modes. This could reveal interesting higher-dimensional structures which emerge in the one-loop four-point Mellin amplitudes of higher KK modes when defects are absent \cite{Aprile:2019rep,Alday:2019nin,Huang:2023ppy,Aprile:2022tzr,Heslop:2023gzr}. 

Second, in the defect-free case an important complementary method is the position space approach \cite{Aprile:2017bgs,Aprile:2017qoy}, which also proved to be useful at higher loops \cite{Huang:2021xws,Drummond:2022dxw,Huang:2023oxf}. A similar approach in position space for defect correlators will be a valuable addition to our toolkit. 

Third, another interesting open question is how to take the flat-space limit for defect correlators. For individual tree-level Witten diagrams, a Mellin space high-energy limit similar to that of \cite{Penedones:2010ue} seems to yield the correct result (see also \cite{Pufu:2023vwo}). However, a rigorous analysis of the flat-space limit is still lacking. It would be desirable to derive such a prescription and study the flat-space limit of our one-loop Mellin amplitude. 

Fourth, it is worth asking about cases with less supersymmetry (or no supersymmetry at all). The essential point, namely the gluing prescription to obtain the LLS, remains unchanged. This is the same as the unitarity method for the defect-free case \cite{Aharony:2016dwx}. The practical difficulty is to solve the mixing problem at tree level, which may involve analyzing a larger set of tree-level correlators as they are no longer related by supersymmetry. On a technical level, one may also encounter additional single poles in the Mellin amplitude due to absence of a reduced correlator (as in e.g. \cite{Alday:2020tgi}).

Finally, in addition to obvious generalizations (Wilson loops, boundaries, etc.), the term ``defect'' encompasses many distinct systems. This includes some setups which are usually not perceived as defects, and hence are drastically different in physics. Here we mention two such examples which recently attracted interest from the integrability community. The first example are correlators in $\mathcal{N}=4$ SYM with two ``giant graviton'' operators (determinant operators) and two light operators \cite{Jiang:2023uut,Brown:2024tru}. The giant gravitons are dual to a D3-brane wrapping S$^3\subset$S$^5$ and the four-point function can be studied as a two-point function with a $0$-dimensional defect \cite{giantgraviton}. The other example is $\mathcal{N}=4$ SYM on $\mathbb{RP}^4$ with gauged charge conjugation, and is dual to a $\mathbb{Z}_2$ quotient of Type IIB string theory on AdS$_5\times$S$^5$ \cite{Wang:2020jgh,Caetano:2022mus}. The $\mathbb{Z}_2$ fixed locus S$^2\subset$S$^5$ is an O1 orientifold and can be viewed as a $(-1)$-dimensional defect. Recently, tree-level correlators have been bootstrapped in \cite{Zhou:2024ekb}. In short, the strategy introduced in this paper is very general and applies to these systems with ``generalized defects'' as well.

\vspace{0.5cm}

\noindent {\bf Acknowledgements.} The work of J.C. and X.Z. is supported by the NSFC Grant No. 12275273, funds from Chinese Academy of Sciences, University of Chinese Academy of Sciences, and the Kavli Institute for Theoretical Sciences. The work of X.Z. is also supported by the NSFC Grant No. 12247103 and the Xiaomi Foundation. H.P. acknowledges support from the FWO grant G094523N. AGG is supported by the Simons Foundation by grants 915279 (IHES) and 733758 (Bootstrap Collaboration).

\appendix

\section{Supplemental Material}

\subsection{Defect CFT kinematics}

We review here the basic kinematics of defect CFTs to fix our conventions (see \cite{Billo:2016cpy} for a more detailed exposition). The bulk one-point and bulk-defect two-point functions are fixed by defect conformal symmetry up to overall OPE coefficients
\begin{equation}
\begin{split}
&\llangle O_{\Delta,\ell,k}(x,v,u) \rrangle=\frac{a_{O}((x^i v^i)^2-|v^i|^2 |x^i|^2)^{\frac{\ell}{2}} (u \cdot \theta)^{k}}{|x^{i}|^{\Delta+\ell}}\;,
\\&\llangle\mathcal{S}_k (x_1, u_1)\, \widehat{O}_{\widehat{\Delta}, s}(\widehat{x}_2, v_2)\rrangle=\frac{b_{k\widehat{O}} (u_1 \cdot \theta)^{k}(x_1^{i}  v_2^i)^s}{|x_1^{i}|^{2 k-\widehat{\Delta}+s} (x_1-\widehat{x}_2)^{2\widehat{\Delta}}}\;,
\end{split}
\end{equation}
where the polarization null vector $v\in\mathbb{R}^6$ is for the 6d spin, and $v_2\in \mathbb{R}^4$ for the transverse spin.

The conformal blocks can be obtained by solving Casimir equations \cite{Billo:2016cpy,Lauria:2017wav,Isachenkov:2018pef,Liendo:2019jpu}.  In the defect channel, the conformal block reads
\begin{equation}
\widehat{f}_{\widehat{\Delta}, s}(z,\bar z)=\frac{(z \bar z)^{\frac{\widehat{\Delta}-s}{2}}}{(1-z \bar z)}\frac{z^{s+1}-\bar{z}^{s+1}}{z-\bar z}\;,
\end{equation}
and in the bulk channel
\begin{equation}
\begin{split}
&f_{\Delta,\ell}^{k_{12}}(z,\bar z)=\frac{(1-z)^2(1-\bar{z})^2 }{(\bar{z}-z)(1-z \bar{z}) (z \bar{z})^{\frac{k_{12}}{2}}}\\{}&\quad\qquad\times\left(k_{\Delta+\ell}(1-z) k_{\Delta-\ell-4}(1-\bar{z})-(z \leftrightarrow \bar{z})\right)\;,
\end{split}
\end{equation}
where $k_{ij}\equiv k_i-k_j$ and
\begin{equation}
k_\beta(z)=z^{\beta / 2}{ }_2 F_1\left(\frac{\beta}{2}-k_{12}, \frac{\beta}{2} ; \beta ; z\right)\;.
\end{equation}

\subsection{Superconformal blocks}

Superconformal symmetry further combines the bosonic conformal blocks into superconformal blocks. Here we collect the results used in the main text where we exchange long superconformal multiplets with R-symmetry singlet super primaries.

Let us first consider the defect channel. The exchanged R-symmetry representations are in the $(r,r)$ representation of $SO(4)=SU(2)\times SU(2)$ and the information is captured by the R-symmetry blocks
\begin{equation}
\widehat{h}_r(\omega)=\frac{(-1)^r}{2^r} \frac{\omega^{r+1}-\omega^{-(r+1)}}{\omega-\omega^{-1}}\;.
\end{equation}
The defect channel superconformal block for defect long multiplets in the singlet representation is a linear combination of contributions of operators in this multiplet
\begin{equation}
\begin{split}
{\bf \widehat{g}}_{\widehat{\Delta},s}=&\;\widehat{f}_{\widehat{\Delta}, s}\widehat{h}_0+2\widehat{f}_{\widehat{\Delta}+1, s-1}\widehat{h}_1+2\widehat{f}_{\widehat{\Delta}+1, s+1}\widehat{h}_1 
  \\&+\widehat{f}_{\widehat{\Delta}+2, s-2}\widehat{h}_0+4\widehat{f}_{\widehat{\Delta}+2, s}\widehat{h}_2+\widehat{f}_{\widehat{\Delta}+2, s}\widehat{h}_0
\\&+  \widehat{f}_{\widehat{\Delta}+2, s+2}\widehat{h}_0+2\widehat{f}_{\widehat{\Delta}+3, s-1}\widehat{h}_1 \\&+2\widehat{f}_{\widehat{\Delta}+3, s+1}\widehat{h}_1+\widehat{f}_{\widehat{\Delta}+4, s}\widehat{h}_0\;,
\end{split}
\end{equation}
where the coefficients are determined by the superconformal Ward identities. This expression is proportional to a single bosonic block with shifted conformal dimension
\begin{equation}\label{fhatshift}
{\bf \widehat{g}}_{\widehat{\Delta},s}=\mathrm{R} \,\widehat{\bf f}_{\widehat{\Delta},s}\;,\quad \widehat{\bf f}_{\widehat{\Delta},s}= \widehat{f}_{\widehat{\Delta}+2,s}\;,
\end{equation}
with the same factor ${\rm R}$ appearing in the solution to the superconformal Ward identities.

In the bulk channel, we restrict to the $k_1=k_2=k$ case. The R-symmetry block for exchanging a rank-$R$ symmetric traceless representation of $SO(5)$ is given by
\begin{equation}
h_R(\sigma)=\sigma^{k-\frac{R}{2}}{ }_2 F_1\left(-\frac{R}{2},-\frac{R}{2} ;-R-\frac{1}{2} ; \frac{\sigma}{2}\right)\;.
\end{equation}
The superconformal block of a bulk-channel R-symmetry singlet long multiplet takes the form
\begin{equation}
\begin{split}
{\bf g}_{\Delta,\ell}=&\;f^0_{\Delta,\ell}h_0+a_1 \,f^0_{\Delta+2,\ell-2}h_0+a_2 \,f^0_{\Delta+2,\ell-2}h_2\\&+a_3\, f^0_{\Delta+2,\ell+2}h_0+a_4 \,f^0_{\Delta+2,\ell+2}h_2\\&+a_5\, f^0_{\Delta+4,\ell-4}h_0+a_{6} \,f^0_{\Delta+4,\ell}h_0\\&+a_{7}\, f^0_{\Delta+4,\ell}h_2 +a_{8}\, f^0_{\Delta+4,\ell}h_4\\&+ a_{9}\, f^0_{\Delta+4,\ell+4}h_0 +a_{10}\, f^0_{\Delta+6,\ell-2}h_0\\&+a_{11} \,f^0_{\Delta+6,\ell-2}h_2+a_{12}\, f^0_{\Delta+6,\ell+2}h_0\\&+a_{13}\, f^0_{\Delta+6,\ell+2}h_2+ a_{14} \,f^0_{\Delta+8,\ell}h_0\;,
\end{split}
\end{equation}
where all the coefficients can be fixed by superconformal Ward identities. The superconformal block can be analogously expressed as a single bosonic conformal block with shifted dimensions
\begin{equation}
\left(\mathcal{D}\mathcal{B}^{-1}\right)^{2k}{\bf g}_{4k+2n+\ell,\ell}=\mathrm{R}\,{\bf f}_{4k+2n+\ell,\ell} \;,
\end{equation}
where
\begin{equation}\label{fshift}
{\bf f}_{4k+2n+\ell,\ell}(z,\bar z,\sigma)=\frac{1}{4}\left(\mathcal{D}\mathcal{B}^{-1}\right)^{2k+2}\sigma^{k-2} f_{4k+4+2n+\ell,\ell}^{0}\;.
\end{equation}

\subsection{A collection of explicit expressions}

Here we collect the explicit expressions for the OPE coefficients which appeared in the main text. The coefficients of the free correlator \eqref{Hfreedefectdec} are
\begin{equation}
\begin{split}
&\langle (b^{(0)}_{k,m,s})^2\rangle=\frac{2^{-k+1}(s+1) (2k+2m+s+1) }{\Gamma (2 k) \Gamma (k-1) \Gamma (k+1)}\\{}&\quad\quad\quad\times\frac{(m+1)_{2 k-1} (m+s+2)_{2 k-1}}{\left((k+m)^2-1\right)\left((k+m+s+1)^2-1\right)}\;.
\end{split}
\end{equation}
The coefficients of the disconnected correlators \eqref{Fdiscbulkdec} are
\begin{equation}
\begin{split}
&\langle\lambda^{(0)}_{k,m,\ell} a^{(0)}_{m,\ell}\rangle=\frac{ 3 \pi\,2^{-2 (2 k+\ell+2 m+3)}  (4 k+\ell+2 m+1)}{\Gamma (2 k) \Gamma (2 k-2)\Gamma (\frac{m}{2}+1)}\\{}&\quad\times\frac{ (\ell+2)\Gamma(k+\frac{m}{2}-1) \Gamma (2 k+m) \Gamma \big(\frac{4 k+m-1}{2}\big)}{ \Gamma(k+\frac{m+3}{2})\Gamma(2 k+m-\frac{1}{2})}\\{}&\quad\times\frac{   \Gamma \big(\frac{2 k+\ell+m}{2}\big) \Gamma (2 k+\ell+m+2) \Gamma \big(\frac{4
   k+\ell+m+1}{2} \big)}{   \Gamma \big(\frac{\ell+m+4}{2}\big) \Gamma (2
   k+\ell+m+\frac{3}{2}) \Gamma \big(\frac{2 k+\ell+m+5}{2}\big)}\;.
\end{split}
\end{equation}
The tree-level defect channel decomposition has coefficients \eqref{H2ktreedefectdec}
\begin{equation}
\begin{split}
 &\langle b^{(0)}_{2,m,s} b^{(0)}_{k,m,s}\widehat{\gamma}^{(1)}_{m,s}\rangle\\{}&= \frac{(-1)^{k+1} (k+m) (k+m+s+1) (m+1)_{2 k-1} }{\Gamma (k-1) \sqrt{2^{k-1}\Gamma (2 k-1)}}\;,
\end{split}
\end{equation}
in terms of which the defect channel one-loop coefficients \eqref{H221loopdefectdec} read
\begin{equation}
\begin{split}
& \frac{1}{2} \langle (b^{(0)}_{2,n,s})^2(\widehat{\gamma}_{n,s}^{(1)})^2\rangle=\frac{L (L+s+1) (L-1)_3 (L+s)_3}{2 (s+1) (2 L+s+1)}\\{}&\times\sum_{k=2}^{L} \frac{(k-1) k (2 k-1) \Gamma (k+L) \Gamma (-k+L+s+2)}{\Gamma (-k+L+1) \Gamma (k+L+s+1)}\;.
\end{split}
\end{equation}
The OPE coefficients of four-point functions can be found in \cite{Alday:2020tgi}. They lead to the following bulk channel coefficients at one loop (c.f. \eqref{oneloopOPEcoeffs})
\begin{equation}
\begin{split}
&\frac{1}{2} \langle\lambda^{(0)}_{2,n,\ell} a^{(0)}_{n,\ell} \gamma^{(1)}_{n,\ell}\rangle=-\frac{\left((2 M-2)^2-1\right)  \left((\ell+2 M)^2-1\right)}{\left((\ell+2)^2-1\right) \left((\ell+4 M+1)^2-1\right)}\\{}&\times\frac{ 2^{-\ell-4 M+3}\Gamma (M) \Gamma (M+2) \Gamma (\frac{\ell}{2}+M+1) \Gamma (\frac{\ell}{2}+M+3)}{ \Gamma(2
   M-\frac{1}{2}) \Gamma(\ell+2 M+\frac{3}{2})}\\{}& \times
\sum_{k=2}^{M}\bigg[\frac{(k-1) k (2 k-1) \Gamma (k+M-\frac{1}{2}) \Gamma(-k+\frac{\ell+3}{2}+M)}{\Gamma (-k+M+1) \Gamma(k+\frac{\ell}{2}+M+1)} \\{}&\quad\quad\quad\times (k (\ell+2) (\ell+4 M+1)+2 M (2 M-1))\Big]\;.
\end{split}
\end{equation}

Let us also give the first few orders of the LLS in an expansion of small cross-ratios. In the defect channel, (\ref{LLSdefect}) can be expanded as
\begin{equation}\label{F12}
\begin{split}
F_{1,2}=&-\tfrac{108}{(\mathcal{B}+1)^{5}}  (\mathcal{B}^4+4 \mathcal{B}^3+6 \mathcal{B}^2+4 \mathcal{B}+2) \mathcal{D}^6\\&- \tfrac{216}{(\mathcal{B}+1)^{7} } (75 \mathcal{B}^6+390 \mathcal{B}^5+826 \mathcal{B}^4\\&+904 \mathcal{B}^3+532 \mathcal{B}^2+96 \mathcal{B}+32)  \mathcal{D}^8\\&-\tfrac{432}{(\mathcal{B}+1)^{9}} (1295 \mathcal{B}^8+8470 \mathcal{B}^7+23835 \mathcal{B}^6\\&+37570 \mathcal{B}^5+36128 \mathcal{B}^4+21492 \mathcal{B}^3\\&+8298 \mathcal{B}^2+972 \mathcal{B}+243) \mathcal{D}^{10}+\mathcal{O}(\mathcal{D}^{12})\;,
\end{split}
\end{equation}
\begin{equation}\label{F02}
\begin{split}
F_{0,2}=&-\tfrac{9}{ (\mathcal{B}+1)^{5}}\big(12 \mathcal{B}^3+42 \mathcal{B}^2+52 \mathcal{B}+25\\&-12 (\mathcal{B}^4+4 \mathcal{B}^3+6 \mathcal{B}^2+4 \mathcal{B}+2) \log (\mathcal{B}+1)\big) \mathcal{D}^6\\&-\tfrac{18}{(\mathcal{B}+1)^{7}} \big(900 \mathcal{B}^5+4230 \mathcal{B}^4+7872 \mathcal{B}^3+7227 \mathcal{B}^2\\&+3274 \mathcal{B}-116-12 (75 \mathcal{B}^6+390 \mathcal{B}^5+826 \mathcal{B}^4\\&+904 \mathcal{B}^3+532 \mathcal{B}^2+96 \mathcal{B}+32) \log (\mathcal{B}+1)\big) \mathcal{D}^8\\& -\tfrac{18}{(\mathcal{B}+1)^{9}} \big(31080 \mathcal{B}^7+187740 \mathcal{B}^6+480760 \mathcal{B}^5\\&+675650 \mathcal{B}^4+562308 \mathcal{B}^3+275298 \mathcal{B}^2\\&+89820 \mathcal{B}-10929-24 (1295 \mathcal{B}^8+8470 \mathcal{B}^7\\&+23835 \mathcal{B}^6+37570 \mathcal{B}^5+36128 \mathcal{B}^4+21492 \mathcal{B}^3\\&+8298 \mathcal{B}^2+972 \mathcal{B}+243) \log (\mathcal{B}+1)\big) \mathcal{D}^{10} \\&+\mathcal{O}(\mathcal{D}^{12})\;.
\end{split}
\end{equation}
In the bulk channel, (\ref{LLSbulk}) can be expanded as 
\begin{equation}\label{F12B}
\begin{split}
F_{1,2}=&\;\tfrac{216 \mathcal{D}^6 \left(6 \mathcal{D}^4+12 \mathcal{D}^2+1\right)}{\left(4 \mathcal{D}^2-1\right)^5}\\
&+\tfrac{216 \mathcal{D}^6 \left(78 \mathcal{D}^4+56 \mathcal{D}^2+3\right)}{\left(4 \mathcal{D}^2-1\right)^6}\mathcal{B}\\
&+\tfrac{432 \mathcal{D}^6 \left(378 \mathcal{D}^6+1245 \mathcal{D}^4+266 \mathcal{D}^2+4\right)}{\left(4 \mathcal{D}^2-1\right)^7}\mathcal{B}^2+\mathcal{O}(\mathcal{B}^3)\;,
\end{split}
\end{equation}
\begin{equation}
\begin{split}\label{F11}
F_{1,1}=&\;\tfrac{18 \mathcal{D}^6}{(4 \mathcal{D}^2-1)^5}  \big(5 \sqrt{1-4 \mathcal{D}^2} (22 \mathcal{D}^2+5)\\&+24(6 \mathcal{D}^4+12 \mathcal{D}^2+1) \log{\mathcal{A}}\big)\\&+ \tfrac{18  \mathcal{D}^6  }{(4 \mathcal{D}^2-1)^6} \big(\sqrt{1-4 \mathcal{D}^2} \left(292 \mathcal{D}^4+734 \mathcal{D}^2+87\right)\\&+24 (78 \mathcal{D}^4+56 \mathcal{D}^2+3) \log \mathcal{A}\big) \mathcal{B}\\&+ \tfrac{36 \mathcal{D}^6  }{(4 \mathcal{D}^2-1)^7} \big(\sqrt{1-4 \mathcal{D}^2} (3541 \mathcal{D}^4+1042 \mathcal{D}^2-29)\\&+24 (378 \mathcal{D}^6+1245 \mathcal{D}^4+266
   \mathcal{D}^2+4) \log{\mathcal{A}}\big) \mathcal{B}^2\\&+\mathcal{O}(\mathcal{B}^3)\;,
\end{split}
\end{equation}
\begin{equation}\label{F10}
\begin{split}
F_{1,0}=&\;\tfrac{9 \mathcal{D}^6 }{(4
   \mathcal{D}^2-1)^5} \big(10 \sqrt{1-4 \mathcal{D}^2} (22 \mathcal{D}^2+5)\log{\mathcal{A}}\\&+24 (6 \mathcal{D}^4+12 \mathcal{D}^2+1)\log^2{\mathcal{A}} \big)\\&+ \tfrac{9  \mathcal{D}^6  }{(4
   \mathcal{D}^2-1)^6} \big(5 (-88 \mathcal{D}^4+2 \mathcal{D}^2+5)\\&+2 \sqrt{1-4 \mathcal{D}^2} (292 \mathcal{D}^4+734 \mathcal{D}^2+87) \log{\mathcal{A}}\\&+24(78 \mathcal{D}^4+56 \mathcal{D}^2+3) \log ^2\mathcal{A}\big) \mathcal{B}\\&+ \tfrac{9 \mathcal{D}^6 }{2 (4 \mathcal{D}^2-1)^7} \big(3 (16384 \mathcal{D}^6+34312 \mathcal{D}^4\\&-5318 \mathcal{D}^2-1071)+8 \sqrt{1-4 \mathcal{D}^2}(3541 \mathcal{D}^4\\&+1042 \mathcal{D}^2-29) \log{\mathcal{A}}\\&+96(378 \mathcal{D}^6+1245 \mathcal{D}^4+266 \mathcal{D}^2+4) \log ^2\mathcal{A}\big) \mathcal{B}^2\\&+\mathcal{O}(\mathcal{B}^3)\;,
\end{split}
\end{equation}
where
\begin{equation}
\mathcal{A}=\frac{2}{1+\sqrt{1-4 \mathcal{D}^2}}\;.
\end{equation}

\bibliography{refs} 
\bibliographystyle{utphys}
\end{document}